\documentclass[useAMS,usenatbib]{mnras}
\usepackage{graphicx}
\usepackage{textcomp}
\usepackage{amssymb}\usepackage{hyperref}

 \usepackage{times}

\newcommand\lsim{\mathrel{\rlap{\lower4pt\hbox{\hskip1pt$\sim$}}
        \raise1pt\hbox{$<$}}}
\newcommand\gsim{\mathrel{\rlap{\lower4pt\hbox{\hskip1pt$\sim$}}
        \raise1pt\hbox{$>$}}}
\newcommand{\be}{\begin{equation}}
\newcommand{\ba}{\begin{eqnarray}}
\newcommand{\ee}{\end{equation}}
\newcommand{\ea}{\end{eqnarray}}

\title[Black hole, neutron star and white dwarf merger rates in AGN disks]{Black hole, neutron star and white dwarf merger rates in AGN disks}

\author[B. McKernan et al.]{B. McKernan$^{1,2,3}$\thanks{E-mail:bmckernan at amnh.org (BMcK)}, K.E.S. Ford$^{1,2,3}$, R.O'Shaughnessy$^{4}$\\
$^{1}$Department of Astrophysics, American Museum of Natural History, New York, NY 10024, USA\\ 
$^{2}$Department of Science, BMCC, City University of New York, New York, NY 10007, USA\\ 
$^{3}$Graduate Center \& CUNY Institute for Astronomy (CIfA)., City University of New York, 365 5th Avenue, New York, NY 10016, USA\\
$^{4}$ Center for Computational Relativity and Gravitation, Rochester Institute of Technology, Rochester, NY 14623, USA\\
}
\begin{document}

\date{Accepted. Received; in original form}

\pagerange{\pageref{firstpage}--\pageref{lastpage}} \pubyear{2008}

\maketitle

\label{firstpage}

\begin{abstract}
Advanced LIGO \& Advanced Virgo are detecting a large number of binary stellar origin black hole (BH) mergers. A promising channel for accelerated BH merger lies in active galactic nucleus (AGN) disks of gas around super-masssive black holes. Here we investigate the relative number of compact object mergers in AGN disk models, including BH, neutron stars (NS) and white dwarfs, via Monte Carlo simulations. We find the number of all merger types in the bulk disk grows $\propto t^{1/3}$ which is driven by the Hill sphere of the more massive merger component. Median mass ratios of NS-BH mergers in AGN disks are  $\tilde{q}=0.07\pm 0.06(0.14\pm 0.07)$ for mass functions (MF) $M^{-1(-2)}$. If a fraction $f_{AGN}$ of the observed rate of BH-BH mergers (${\cal}{R}_{BH-BH}$) come from AGN, the rate of NS-BH (NS-NS) mergers in the AGN channel is ${\cal}{R}_{BH-NS} \sim f_{AGN}[10,300]\rm{Gpc}^{-3} \rm{yr}^{-1}$,(${\cal}{R}_{NS-NS} \leq f_{AGN}400\rm{Gpc}^{-3} \rm{yr}^{-1}$). Given the ratio of NS-NS/BH-BH LIGO search volumes, from preliminary O3 results the AGN channel is not the dominant contribution to observed NS-NS mergers. The number of lower mass gap events expected is a strong function of the nuclear MF and mass segregation efficiency. Compact object merger ratios derived from LIGO can restrict models of MF, mass segregation and populations embedded in AGN disks. The expected number of EM counterparts to NS-BH mergers in AGN disks at $z<1$ is $\sim [30,900]{\rm{yr}}^{-1}(f_{AGN}/0.1)$.  EM searches for flaring events in large AGN surveys will complement LIGO constraints on AGN models and the embedded populations that must live in them.
\end{abstract}


%%%%%%%%%%%%%%%%%%%%%%%%%%%%%%%
\section{Introduction}
Advanced LIGO \citep{Aasi15} and Advanced Virgo \citep{Acernese15} are detecting a surprisingly large rate of stellar mass black hole (BH) binary mergers \citep{Abbott18GWTC1}.  While a joint analysis of the observed population suggests most merging binaries total mass is  less than $45 M_\odot$ \citep{Abbott18Pop}, it is possible that high- or marginal-significance merger events, if of astrophysical origin, may have BH masses above the pair-instability threshold \citep{Woosley17}, as detected by LIGO in events GW170929 \citep{KC2019}, GW170817A \citep{Zackay19},  and IMBHC-170502 \citep{Udall19}.
Such BH masses imply a hierarchical merger origin in a deep-potential galactic nucleus \citep{Gerosa19}. In this environment an AGN disk is preferred as a merger accelerant \citep{McK12,McK14,Bellovary16,Bartos17,Stone17,McK18,Secunda18,Yang19,McK19a} and as an inhibitor of binary disruption via dynamically 'hot' tertiary encounters \citep{Leigh18}. Thus, a handful of LIGO-detected events in the \emph{upper} mass gap, expected from the supernova pulsational pair-instability, implies a large number of lower mass BH-BH mergers in this same channel \citep{Gerosa19}. Following this line of reasoning implies it is not unreasonable to expect about half of the observed LIGO BH-BH events are happening in AGN disks.

BH-BH mergers are not the only mergers that will occur in AGN disks. A nuclear star cluster will contain neutron stars (NS), white dwarfs (WD) as well as main-sequence and evolved stars and some fraction of all of these populations must also end up in AGN disks \citep{McK12}. 
These nuclear cluster objects will migrate, form binaries and merge within AGN disks. As a result, muffled electromagnetic (EM) signatures due to e.g. tidal disruptions of NS by BH; kilonovae from NS-NS mergers and supernovae from WD mergers or tidal disruptions might be detectable in large surveys of AGN \citep{Graham17,Cannizzaro19}. We should also expect merged products of lower mass objects (NS,WD) within the \emph{lower} mass gap, possibly like the merger event GW190425z \citep{LIGO20}.

Here we extend our investigations in \citet{McK19a} to study Monte Carlo simulations of mergers in populations of compact objects, including BH, NS (neutron stars) and WD (white dwarfs) in AGN disks. We consider several models of the initial disk population and compare relative merger rates among and between the different compact object populations. Such rates allow us to constrain both the AGN channel events detected in GW by Advanced LIGO and Virgo, as well as the rate of occurrence of tidal disruption events and Type Ia supernovae detectable in EM bands in AGN disks \citep[e.g.][]{Graham17,Cannizzaro19}.

\section{Compact object populations in galactic nuclei}
We expect a large population of compact objects in galactic nuclei as a function of stellar evolution, dynamical friction and a deep potential well around the SMBH \citep[e.g.][]{Morris93, Miralda00}. Galactic nuclei containing small mass SMBH ($<$ few $\times 10^{7}M_{\odot}$) often have associated dense nuclear star clusters (NSC) \citep{Graham09}. Galactic nuclei with more massive SMBH can contain NSCs but the NSC mass as a fraction of spheroid mass appears to decrease.
In our own Galaxy the mass function ($M^{-\Gamma}$) of the NSC is flatter than a Kroupa/Salpeter MF  \citep{Lu13}, with $95\%$ confidence interval spanning $\Gamma=[1.1,2.1]$ for NSC age $\geq 3.3$Myr and $\Gamma=[1.5,2.1]$ if NSC age $<3.3$Myr. A flatter than normal $\Gamma$ may be suggestive of mass segregation or a top-heavy IMF for star formation. Also suggestive of segregation are observations of low mass stars ($\sim 1M_{\odot}$) in our Galactic nucleus revealing a flatter spatial density profile ($\propto r^{-[1.1,1.4]}$) than expected for a Bahcall-Wolf ($r^{-7/4}$) single model profile \citep{Schodel18}. There is also a considerable  population of BH X-ray binaries in our Galactic nucleus \citep{Hailey18}. Thus, a much larger ($\sim 10^{4}$ BH) underlying compact object (CO) population of singletons is inferred in the central $\rm{pc}^{3}$ \citep{Generozov18}, representing the densest population of BH in the local Universe. 

A Milky-Way like nuclear star cluster formed by the mergers of globular clusters over time implies different density profiles for COs and stars. BH dominate the mass density inside small distances ($<0.3$pc) due to mass segregation and energy equipartition \citep[e.g.][]{Antonini14,Baumgardt19}.
Around smaller mass SMBH ($< \rm{few} \times 10^{7}M_{\odot}$), dense nuclear star clusters are observed and the timescale for mass segregation is short. In such circumstances, we might find a BH-dominated mass function embedded in the disk. Around more massive SMBH, mass segregation may be less efficient and a less top-heavy IMF may be more appropriate. 
New CO formation might come from extremely top-heavy IMF ($\sim M^{-0.5}$) disky star formation \citep{Bartko10}. \citet{Generozov18} assume a rate of NS deposition in the galactic nucleus $\times 2$ the BH rate. After 10Gyr they  find a slightly shallower density profile for NS than BH, consistent with expectations for a two-component mass  model \citep{Alexander09}. The NS number density within $r<1$pc is within a factor $\sim 2$ similar to the BH number density. 

In the simulations below we choose an IMF ($M^{-\Gamma}$, $\Gamma=1,2$) for the BH, matching the range of observations for the high mass stars in our own NSC. For the added WD and NS populations, we assumed either a model of strong mass segregation or no mass segregation. This allows us to illustrate likely limits to the relative merger numbers. Our no mass segregation model simply assumes a continuation of the powerlaw IMF (when $\Gamma=1,2$) to lower WD, NS masses. Our strong mass segregation model assumes a  broken powerlaw IMF with $\Gamma=0$ for all masses less than the smallest BH mass ($M<5M_{\odot}$) with $\Gamma=1,2$ for the BH IMF. Separately we test the impact of the lower mass gap by reducing the lowest BH mass from $5M_{\odot}$ to $3M_{\odot}$. 
We assumed that the NS and WD accrete from the gas disk at the Eddington rate. For simplicity we also approximated the merged mass of NS and WD encounters with the same approximation we used for BH mergers \citep{Tichy08}, which we expect is accurate to order of magnitude, at least for BH-NS mergers. 

\section{Compact object mergers in AGN disks}
\subsection{Changes to the simulations in \citet{McK19a}}
For extensive discussion of the many assumptions and caveats in the models underlying our simulations, see \citet{McK19a}. In brief, here we consider populations of NSC stellar remnants (BH, NS, WD), embedded in a model AGN disk, undergoing Type I migration, forming binaries and merging. We use two different AGN disk models. Neither model is fully self-consistent, but the \citet{SG03} model, derived from spectral fits to the blue-UV spectra of AGN disks, is more plausible for inner disk properties ($\leq 10^{3}r_{g}$). The \citet{Thompson05} model, derived from mass-inflow from nuclear star formation, is likely more plausible for outer disks ($>10^{3}r_{g}$). Since Type I migration timescales \citep{Tanaka02,Paar10} are radial functions of the disk aspect ratio and surface density, we evaluate the migration timescale as a function of migrator radial position in the disk in our simulations. From 3-d N-body simulations, \citep{Secunda18,Secunda20} we find that as long as the migration timescale at a given radius is longer than the orbital timescale (generally true across all our disk models), binaries form once the binary binding energy inside its mutual Hill sphere is greater than the relative kinetic energy of the two masses. In these simulations we assume that once the binary lies within its mutual Hill sphere this condition is satisfied (for the most massive pair of objects within that Hill sphere), which generally agrees with the tests in \citep{Secunda20}. Since the mass ratio of the most massive IMBH produced in any of our simulations to the SMBH is $q \sim O(10^{-5})$ we ignore the effect of gap-opening as IMBH grow at the migration trap. In \citet{McK19a} we ran simulations R1-R12 that studied the variation of: BH mass function(R1-R3:$M^{-1}$; R4-R12: $M^{-2}$), BH IMF upper limit ($15M_{\odot}$ in R6, otherwise $50M_{\odot}$) BH spin distribution (R7: (1-a), otherwise uniform), AGN lifetime ($5$Myrs in R5,R9,R10, otherwise $1$Myr), the absence of migration traps (R11), inefficient disk capture from the spherical component of the NSC (R8-R10), as well as retrograde  binary hardening efficiency (R2,R10). We allow accretion onto embedded objects to change the initial spin and torque the spin component into alignment with the AGN disk over time and we evaluate the distributions of merger masses, mass ratios, $\chi_{\rm eff}$. 

For the simulations discussed here we made some minor modifications. To the input BH mass function we added delta functions at mass $0.6M_{\odot}$ and $1.3M_{\odot}$ respectively. For simplicity we assume a single uniform value for the initial WD and NS masses, rather than a distribution of masses. The NS and WD delta functions have the same effective widths as the initial BH mass bins (i.e. $\pm 0.5M_{\odot}$). The normalization of the WD, NS delta functions in our simulations depended on the efficiency of mass segregation we assumed. Where the mass segregation in a galactic nucleus is assumed efficient the WD, NS delta-functions have the same normalization as the lowest mass BH (at $5M_{\odot}$), corresponding to a $M^{0}$ low mass MF component. To test circumstances where mass segregation is inefficient, the delta-function normalization corresponded to an  interpolation of the BH IMF to NS ($1.3M_{\odot}$) and WD ($0.6M_{\odot}$) masses. In this work we are most interested in the relative rates of merger of different species(BH,NS,WD), so allowing a mass function for WD, NS should not make a significant difference to the relative merger rates. Where multiple migrators ended up within a mutual Hill sphere, we assumed that (unmodelled) dynamics would always result in the formation of the most massive possible binary. We also test the effect of a lower mass gap by changing the lowest BH mass from $5M_{\odot}$ to $3M_{\odot}$.

\subsection{$M^{-2}$ IMF}
We selected as a baseline model, run R4, from \citet{McK19a} for a  $M^{-2}$ IMF model. For a full description of the caveats, simplifications and prescriptions underlying this run see the discussion in \citet{McK19a}. We modified run R4 by including additional populations of neutron stars (NS) of mass $M_{\rm NS}=1.3M_{\odot}$ and white dwarfs (WD) of mass $M_{\rm WD}=0.6M_{\odot}$ as delta functions in our mass function (MF). Modified run R4 corresponds to a strong mass segregation scenario where the number of WD and NS matches the number of BH of mass $5M_{\odot}$. For our model of strong mass segregation in a NSC, we modified this run (R4) by adding a population of WD, NS with a $M^{0}$ mass distribution that joins the BH $M^{-2}$ IMF. Run R4r is a no mass segregation case, where we reduced the normalization on the BH IMF in run R4 and the WD, NS delta-function normalizations were interpolated from the BH IMF. 

We tested variations in model assumptions by adding the same normalization low mass delta-functions for WD,NS to the BH IMF in runs R6,R8, R11 and R12 from \citet{McK19a}. This allowed us to test: a change the range of the BH IMF (R6), the absence of ground-down orbits (R8), the absence of a migration trap (R11) and a different disk model (R12) \citep{Thompson05}. 

\begin{figure}
\begin{center}
\includegraphics[width=6.0cm,angle=-90]{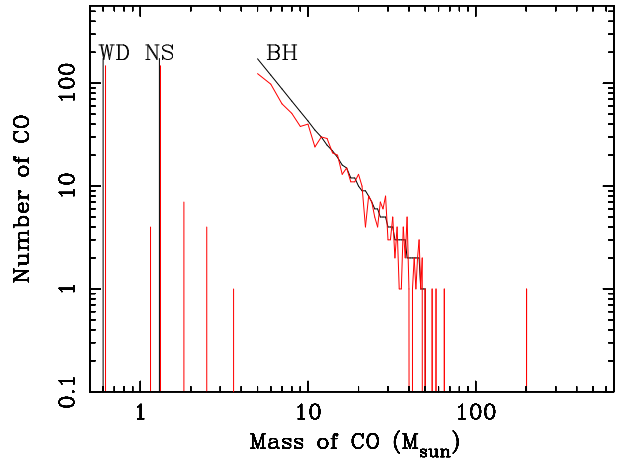}
\end{center}
\caption[Mass spectrum]{Evolution of mass spectrum of BH and NS in the disk as a result of mergers and gas accretion for a single modified run of R4. Black solid lines correspond to the input initial mass function (IMF). Red solid line corresponds to the mass distribution after 1Myr. Ground-down objects added to the disk over time are drawn from a distribution identical to the IMF.
\label{fig:mass_spectrum}}
\end{figure}

Fig.~\ref{fig:mass_spectrum} illustrates the results of a single run of our broken powerlaw model corresponding to strong mass segregation. The mass function changes from the IMF (black curve) to the final mass function (red curve) after 1Myr of migration, collisions and binary evolution. Merged WD and merged NS are apparent, along with a $\sim 200M_{\odot}$ IMBH at the migration trap. Objects on ground-down orbits are selected from the IMF and support the original mass function over time.

\begin{table*}
 \begin{minipage}{130mm}
   \caption{Number of merger types from 100 runs with a broken powerlaw IMF of $\Gamma=0$ spanning WD, NS and the lowest mass BH, and $\Gamma=2$ for the BH mass function. Column 1 is the run number. Column 2 is the powerlaw index either side of the lowest mass BH. Column 3 is the number of BH-BH mergers ($N_{\rm BB}$). Column 4 is the number of BH-NS mergers expressed as a fraction of the BH-BH number. Columns 5,6 are as column 4 except for BH-WD and BH-X mergers (where X is less massive than a BH, but more massive than a NS, e.g. X= result of NS-NS merger or NS-WD merger etc.). Columns 7,8 and 9 are as column 4 but for NS-NS, WD-WD and NS-WD mergers respectively. Column 10 is the maximum mass BH in the disk. Column 11 is the median mass involved in mergers at the trap.
    \label{tab:r4bx100}}
\begin{tabular}{@{}lcccccccccc@{}} 
\hline
Run & $\Gamma$ &$N_{\rm BB}$ & $f_{\rm BN}$& $f_{\rm BW}$  &$f_{\rm BX}$ & $f_{\rm NN}$ & $f_{\rm WW}$& $f_{\rm NW}$ & $M_{\rm max,bulk}$& $\tilde{M}_{\rm trap}$\\

\hline

R4 & 0,2 &$12917$ & $0.29$ & $0.28$ & $0.01$ & $0.017$ & $0.013$&$0.035$ & $105.6$& $97.6 \pm 71.7$  \\

R6 & 0,2 &$12077$ & $0.39$ & $0.36$ & $0.01$ & $0.03$ & $0.02$&$0.05$ &$103.8$ & $68.1 \pm 40.4$\\

R8 & 0,2 &$12982$ & $0.38$ & $0.29$ & $0.008$ & $0.016$&$0.014$ & $0.033$ &$143.4$ & $91.3 \pm 67.4$\\

R11 & 0,2 &$13059$ & $0.29$ & $0.28$ & $0.01$ & $0.017$ & $0.01$&$0.032$ & $108.0$ & -\\

R12 & 0,2 &$2241$ & $0.23$ & $0.23$ &$0.004$  & $0.023$ &$0.017$ &$0.045$ & $147.4$ & $719.7 \pm 444.7$ \\
 
 \hline
R4r & 2,2 &$185$ & $3.18$ & $15.29$ & $1.7$ & $3.7$ & $63.7$&$31.7$ & $116.8$& $1.3^{+17.3}_{-0.7}$  \\
R4l & 0,2 & $11742$ &$0.44$ &$0.43$ &$0.02$ &$0.042$ &$0.034$ &$0.08$ & $111.8$ & $56.2 \pm 47.7$ \\
\hline
\end{tabular}
\end{minipage}
\end{table*}

Table~\ref{tab:r4bx100} shows the results from Monte Carlo simulations of runs with a broken powerlaw IMF (runs R4-R12), corresponding to a model of strong mass segregation and for a single powerlaw IMF (R4r), corresponding to a model of no mass segregation, together with run R4l which tests the effect of the extent of the lower mass gap (see \S\ref{sec:lower} below). The relative numbers of species (BH, NS, WD) seem to drive the results. This is because mergers are dominated by the most numerous species: WD for the single powerlaw IMF, BH for the broken powerlaw IMF respectively.

\begin{figure}
\begin{center}
\includegraphics[width=6.0cm,angle=-90]{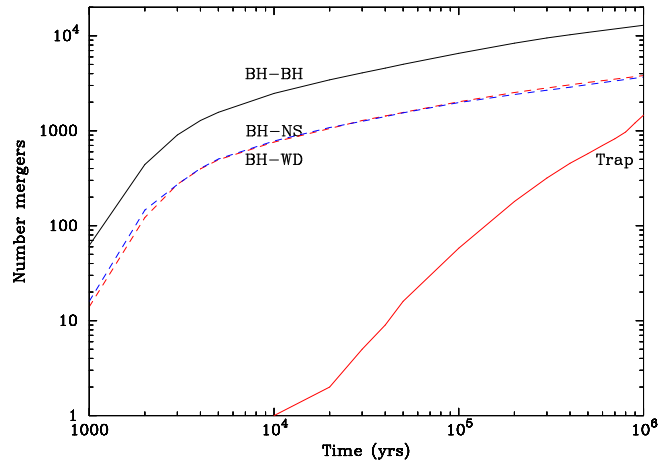}
\end{center}
\caption[Mergers]{Evolution of merger number over time for 100 runs of R4. The number of BH-BH, BH-NS and BH-WD mergers in the bulk of the disk grows approximately as $\propto t^{1/3}$ after $\sim 0.01$Myr. The merger number grows at the trap as approximately $\propto t^{3/2}$. 
\label{fig:mergers}}
\end{figure} 

There is a large difference in the merger ratios depending on whether we include mass segregation or not. If there is no mass segregation (run R4r), the merger rates of the most numerous species (WD) dominates. In R4r WD-WD mergers are most common, followed by WD-NS and then WD-BH. With the addition of strong mass segregation, this picture changes dramatically. For example, in the case of runs R6 and R12, BH are the most numerous species and the BH merger rates dominate. Following  \citet{Dittmann20} we ran a variant of R12 without an explicit migration trap. In this case we found that a large single IMBH does not form. Rather migration seems to become less efficient around $\sim 300r_g$ and a number of IMBH seeds (typically in the range $60-200M_{\odot}$) tend to form and build up in the inner disk. While we ignore tertiary encounters/resonances in these simulations, we expect that as in-migrating lower mass objects continue to reach this region of the inner disk, they will tend to perturb the orbits of IMBH seeds and promote ongoing merger, rather like the scenario in \citet{Secunda20}, although this requires further study.

Fig.~\ref{fig:mergers} shows the change in merger number over time for run R4. There is a burst of mergers early on as objects located in close proximity merge quickly. Although we assume an initial population of singletons in the disk, this early merger burst effectively corresponds to the gas-driven merger of a initial binary fraction $\sim 10\%$ due to random placement within the disk. The initial expected binary fraction in NSCs can range from a few percent up to $20\%$ depending on assumptions \citep{Antonini14}, so we regard this initial burst as an acceptable initial feature of the random initial placements in the simulations. Behaviour after $0.01$Myrs in the simulations is more representative of the general pattern of migrators finding new partners and merging. After $\sim 0.01$Myr, the merger numbers of BH-BH and BH-NS,BH-WD grow at similar rates ($\propto t^{1/3}$). Mergers start at the migration trap around $\sim 0.01$Myr and grow at a much faster rate ($\propto t^{3/2}$).

\begin{table}
 \begin{minipage}{90mm}
   \caption{Median mass ratios and standard deviation from different merger types in Table~\ref{tab:r4bx100}. 
    \label{tab:r4bxq}}
\begin{tabular}{@{}lcccc@{}} 
\hline
Run   & $\tilde{q}_{\rm BB}$& $\tilde{q}_{\rm BN}$& $\tilde{q}_{\rm BW}$& $\tilde{q}_{\rm BX}$\\
\hline

R4 & $0.55\pm 0.25$  & $0.14\pm 0.07$ &$0.06\pm 0.03$ & $0.18\pm0.12$\\
R4l & $0.50 \pm 0.27$& $0.22 \pm 0.14$&$0.10\pm 0.06$&$0.35 \pm 0.22$ \\
R6 & $0.67\pm 0.20$  & $0.18\pm 0.06$ &$0.08\pm 0.03$ & $0.26 \pm 0.11$\\
R8 & $0.55\pm 0.25$  & $0.13\pm 0.07$ &$0.06 \pm 0.04$ & $0.20\pm0.12$\\
R11 & $0.55\pm 0.25$  & $0.14\pm 0.08$ &$0.06\pm 0.04$ & $0.28\pm 0.13$\\
R12 & $0.61\pm 0.26$  & $0.16\pm 0.07$ &$0.07 \pm 0.03$ & $0.16\pm 0.06$\\

\hline
\end{tabular}
\end{minipage}
\end{table}

\begin{table}
 \begin{minipage}{90mm}
   \caption{Median mass ratios and standard deviation from different merger types in Table~\ref{tab:r4bx100}. Column 1 is the run number. Column 2 is the number of BH-BH mergers ($N_{\rm BB}$). Column 3 is the number of BH-NS mergers expressed as a fraction of the BH-BH number. Columns 4,5 are as column 3 except for BH-WD and BH-X mergers (where X is less massive than a BH, but more massive than a NS, e.g. X= result of NS-NS merger or NS-WD merger etc.). 
    \label{tab:r4bxq}}
\begin{tabular}{@{}lcccc@{}} 
\hline
Run   & $\tilde{q}_{\rm BB}$& $\tilde{q}_{\rm BN}$& $\tilde{q}_{\rm BW}$& $\tilde{q}_{\rm BX}$\\
\hline
R1a & $0.48\pm 0.25$  & $0.07^{+0.07}_{-0.06}$ &$0.03^{+0.03}_{-0.02}$ & $0.18\pm 0.12$\\
R1b & $0.50\pm 0.26$  & $0.068\pm 0.067$ &$0.032\pm 0.031$ & $0.09^{+0.10}_{-0.07}$ \\
R1c & $0.50\pm 0.26$  & $0.07\pm 0.068$ &$0.0316\pm 0.031$ & $0.10 \pm 0.09$ \\
\hline
\end{tabular}
\end{minipage}
\end{table}

\subsection{$M^{-1}$ IMF}
For our $M^{-1}$ IMF we chose as our baseline, run R1 from \citet{McK19a}. We modified run R1 by including additional populations of neutron stars (NS) of mass $M_{\rm NS}=1.3M_{\odot}$ and white dwarfs (WD) of mass $M_{\rm WD}=0.6M_{\odot}$. Run R1a is a strong mass segregation scenario where the MF is flat ($M^{0}$) out to the smallest BH mass ($5M_{\odot}$) and $M^{-1}$ thereafter, the grind-down population mimics the IMF and the disk lasts for $1$Myr. Run 1b is a no mass segregation case where we simply interpolated the $M^{-1}$ IMF down to the WD mass, the grind-down population mimics the IMF and the disk lasts for $1$Myr. Run 1c is as run 1b but we extended the disk lifetime to $5$Myrs.

Table~\ref{tab:r1bx100} shows the results of 100 runs of R1a,R1b and R1c with different seeds for each run. Fig.~\ref{fig:norm_rates_r1} plots the merger number expressed as a fraction of the BB merger number. From Fig.~\ref{fig:norm_rates_r1} the suppression of the low mass population makes a clear difference in the relative rates of BH-NS and BH-WD mergers. The BH-NS and BH-WD fraction of mergers drop by factors of $\sim 4$ and $\sim 10$ respectively for our model of strong mass segregation. 

BH mergers with objects in the lower mass gap (denoted by BX in Table~\ref{tab:r1bx100} and Fig.~\ref{fig:norm_rates_r1}) correspond to a few percent of the BH-BH merger number in R1b. As we extend the disk lifetime in R1c, the BX merger number increases to nearly $10\%$ of the BH-BH merger number. This is the only significant change in the relative fractions of mergers with disk age, since the number of WD, NS merger products increases with disk age and those can go on to merger with more rapidly in-migrating BH. If we assume no mass segregation in the nuclear star cluster and have a single powerlaw IMF with $\Gamma=1$, as in R1c, the BX merger number drops by an order of magnitude. The WD-NS merger number is consistently higher than the NS-NS and WD-WD merger number in all of our simulations. 

\begin{table*}
 \begin{minipage}{145mm}
   \caption{Number of merger types from 100 runs with BH IMF of $\Gamma=1$ and initial conditions from run R1 in \citet{McK19a}. Run R1a corresponds to strong mass segregration in the galactic nucleus with $N_{\rm NS},N_{\rm WD}=N_{\rm BH}(5M_{\odot})$, or a broken powerlaw IMF with $\Gamma=0,1$. Run R1b and run R1c correspond to no mass segregation and a single powerlaw IMF with $\Gamma=1$ and a disk lifetime of 1(5)Myr respectively. Initial conditions for each run were as in \citet{McK19a} and $N_{\rm gr}$ scaled appropriately for the total number of COs in run.  Column 1 is the run number. Column 2 is the IMF powerlaw either side of the break at lowest BH mass. Column 3 is the disk lifetime ($\tau_{\rm AGN}$). Column 4 is the number of BH-BH mergers ($N_{\rm BB}$). Column 5 is the number of BH-NS mergers expressed as a fraction of the BH-BH number. Columns 6,7 are as column 4 except for BH-WD and BH-X mergers (where X is less massive than a BH, but more massive than a NS, e.g. X= result of NS-NS merger or NS-WD merger etc.). Columns 8,9 and 10 are as column 5 but for NS-NS, WD-WD and NS-WD mergers respectively. Column 11 is the maximum mass BH in the disk. Column 12 is the median mass involved in mergers at the trap.
    \label{tab:r1bx100}}
\begin{tabular}{@{}lcccccccccccc@{}} 
\hline
Run & $\Gamma$& $\tau_{\rm AGN}$ & $N_{\rm BB}$ &$f_{\rm BN}$& $f_{\rm BW}$  &$f_{\rm BX}$ & $f_{\rm NN}$ & $f_{\rm WW}$& $f_{\rm NW}$ & $M_{\rm max,bulk}$& $\tilde{M}_{\rm trap}$\\

\hline
R1a & 0,1 & $1\rm{Myr}$ & $15312$ &$0.12$ & $0.12$ & $0.002$ & $0.002$ & $0.002$&$0.005$ &$151.1$ & $216.9 \pm 146.2$\\
R1b & 1,1 & $1{\rm Myr}$& $12954$ & $0.43$ & $0.98$ & $0.03$ & $0.03$ & $0.13$ & $0.15$ & $181.9$ & $154.5 \pm 106.4$\\
R1c & 1,1&$5{\rm Myr}$ & $16908$& $0.44$ & $1.0$ & $0.07$ & $0.03$ & $0.12$&$0.14$ &$219.9$ & $1731.0 \pm 1108.8$\\

\hline
\end{tabular}
\end{minipage}
\end{table*}

\subsection{Testing the lower mass gap}
\label{sec:lower}
In the results outlined above we limited our lowest BH mass to $5M_{\odot}$. We tested the effect of choice of lower bound BH by re-running run R4 with a new BH lower bound of $3M_{\odot}$ as run R4l in Table~\ref{tab:r4bx100} and Table~\ref{tab:r4bxq}. We used the mass segregation model to fix the number of WD and NS to the number of BH at $3M_{\odot}$. As a result we find that mergers involving at least one non-BH increased by a factor of 2.  The median mass ratio in BH-BH mergers decreases slightly from $q=0.55$ to $q=0.50$, as we should expect by increasing the effective mass range that we are selecting mergers from. The median mass ratio in BH-NS, BH-WD, BH-X mergers also increases, which is consistent with a larger number of lower mass BH available for merger. E.g. BH-NS goes from median $\tilde{q}= 0.14$ to median $\tilde{q}=0.22$, with a larger associated variance. Likewise the median mass merging at the migration trap also drops, reflecting the typically lower mass mergers in the bulk disk. Thus  the distribution of mass ratios of mergers from the AGN channel may provide a constraint on the presence or absence and extent of a lower mass gap.

\section{Discussion}
\subsection{Relative merger numbers}
\label{sec:rel}
Fig.~\ref{fig:mergers} shows the evolution of the merger number over time in run R4 in Table~\ref{tab:r4bx100}. After the rapid initial burst of mergers at $<0.01$Myr, the number of BH-BH, BH-NS and BH-WD mergers appears to grow approximately as $t^{1/3}$. Migration timescales are $\propto M^{-1}$, which means the distance migrated within the disk is $\propto M$. The (Hill sphere) scaling for binary formation is $\propto M^{1/3}$. So the merger number evolution as $\propto t^{1/3}$ after the initial burst of mergers, seems to be mostly driven by the more massive in-migrator (the BH), catching up with and merging with less massive (slower) targets.

We write the expected encounter number $N_{i,j}$ between species $i,j$ of median mass $\tilde{M}_{i}\geq \tilde{M}_{j}$ and number density of median mass species members $N_{i},N_{j}$ approximately  as
\begin{equation}
    N_{i,j} \propto N_{i}N_{j} \left(\tilde{M}_{j}\tilde{M}_{i}\right)^{1/3}.
\end{equation}
Thus, the relative median numbers of mergers in this toy model goes generally as
\begin{equation}
    \frac{N_{i,j}}{N_{k,l}} = C \frac{N_{i}N_{j}}{N_{k}N_{l}}\left(\frac{\tilde{M}_{i}\tilde{M}_{j}}{\tilde{M}_{k}\tilde{M}_{l}}\right)^{1/3}
\label{eq:rel_general}
\end{equation}
where $C$ is a proportionality constant of order a few. Thus, for example:
\begin{equation}
    \frac{N_{BN}}{N_{BB}} = C \frac{N_{N}}{N_{B}}\left(\frac{\tilde{M}_{N}}{\tilde{M}_{B}}\right)^{1/3}
    \label{eq:rel}
\end{equation}
and
\begin{equation}
    \frac{N_{WW}}{N_{BB}} = C \left(\frac{N_{W}}{N_{B}}\right)^{2}\left(\frac{\tilde{M}_{W}}{\tilde{M}_{B}}\right)^{2/3}
    \label{eq:rel2}
\end{equation}
where $N_{W},N_{N},N_{B}$ are the number of WD,NS and BH in the disk and $\tilde{M}_{W},\tilde{M}_{N},\tilde{M}_{B}$ are their respective median masses. Parameterizing eqn.~(\ref{eq:rel_general}) for BH, NS and WD in the broken (single) powerlaw $M^{-2}$ IMF runs, we find $C \sim 3(1.5)$ gives the best fit.

\begin{figure}
\begin{center}
\includegraphics[width=6.0cm,angle=-90]{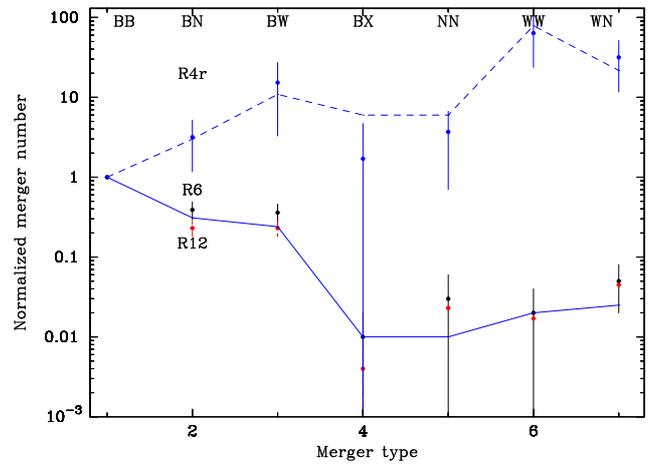}
\end{center}
\caption[Rates]{Merger type expressed as a fraction of the BH-BH (BB) merger number for representative runs listed in Table~\ref{tab:r4bx100}. Results are for R4r (blue points), R6 (black points) and R12 (red points).  Merger types are 1)BH-BH (BB), 2) BH-NS (BN), 3) BH-WD (BW), 4) NS-NS (NN), 5) WD-WD (WW), 6) WD-NS (WN), 7) BH-X (BX) where X is a merger product with mass less than BH. The plot can be read as, e.g. the BH-NS merger number as a fraction of BH-BH merger number for these models spans $\sim 0.2$ (R12), $\sim 0.4$ (R6) and $\sim 3$ (R4r). Dashed and solid lines correspond to simple analytic model fits (see text). Run-to-run variation of the fraction is indicated by the error bars. R4r has larger variance between runs because of the smaller total number of particles.
\label{fig:norm_rates}}
\end{figure}

Fig.~\ref{fig:norm_rates} plots merger numbers of different species normalized to the BH-BH merger number for representative runs from Table~\ref{tab:r4bx100}. Dashed lines correspond to our toy analytic model eqn.~(\ref{eq:rel_general}) and fit the results fairly well. We assumed median masses $\tilde{M}_{BH}=10M_{\odot}$, $M_{NS}=1.3M_{\odot}, M_{WD}=0.6M_{\odot}, M_{X}=1.15M_{\odot}$ in eqn.~(\ref{eq:rel_general}). From Fig.~\ref{fig:mergers} mergers at very early times ($<0.005$Myr) are mostly a population of randomly formed initial binaries merging, not from migration and capture, so we exclude these mergers. As a result, the fractional rates plotted differ slightly from those in Table~\ref{tab:r4bx100}.  The NN and BX merger fractions are least well fit by our toy model. In part this is probably because N and X are usually the smallest species numbers in our simulations. An added complication is that merger products (X) are only produced in large numbers as time progresses, so using their total number in eqn.~(\ref{eq:rel_general}) is a poor approximation. In spite of these complications, our simplistic, few-parameter, analytic model (eqn.~\ref{eq:rel_general}) is a surprisingly reasonable zeroth-order approximation to the results of our simulations.

\begin{figure}
\begin{center}
\includegraphics[width=6.0cm,angle=-90]{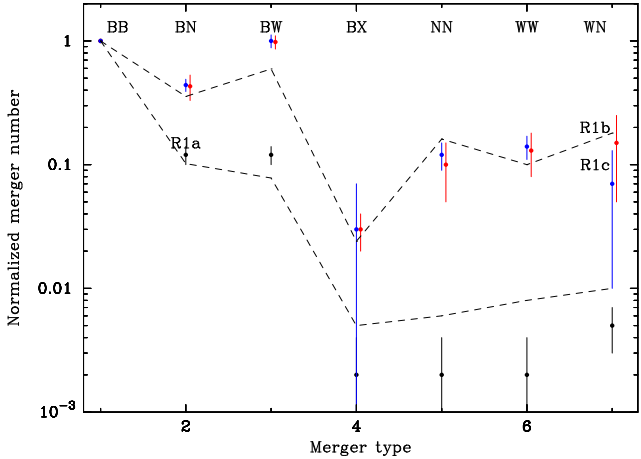}
\end{center}
\caption[Rates]{Number of mergers of a particular type plotted as a fraction of the BH-BH merger number from Table~\ref{tab:r1bx100}. Results are for R1a (solid black line), R1b (solid red line), R1c (solid  blue line). Merger types are 1)BH-BH, 2) BH-NS, 3) BH-WD, 4) BH-X, 5) NS-NS, 6) WD-WD, 7) WD-NS, where X is a merger product with mass greater than NS but less than BH. Dashed lines correspond to simple analytic model fits (see text). Run-to-run variation of the fraction is indicated by the error bars.
\label{fig:norm_rates_r1}}
\end{figure}

Fig.~\ref{fig:norm_rates_r1} is as Fig.~\ref{fig:norm_rates} except for runs from Table~\ref{tab:r1bx100}, excluding mergers at $<0.005$Myr. Dashed lines correspond to our simple analytic model fit from eqn.~(\ref{eq:rel_general}). Our analytic model is generally a poorer fit here than in Fig.~\ref{fig:norm_rates}. Part of the reason for poorer fits to mergers involving BH may be the bigger standard deviation in the merging BH mass distribution with a $M^{-1}$ IMF. 

\subsection{Implications for LIGO}
Let us define the maximum detectability distance for LIGO to BH-BH, BH-NS and BH-X mergers respectively as $D_{L,BB},D_{L,BN},D_{L,BX}$. We can then write the relative numbers of  such mergers detected by LIGO ($N_{L,Bi}/N_{L,BB}$) as
\begin{equation}
    \frac{N_{L,Bi}}{N_{L,BB}}=\frac{N_{Bi}}{N_{BB}} \left(\frac{D_{L,Bi}}{D_{L,BB}}\right) ^{1/3}
\end{equation}
where $i=N,W,X$.  Using the standard approximation $D_L\propto {\cal M}^{5/6}$ where ${\cal M}=(m_1 m_2)^{3/5}/(m_1+m_2)^{1/5}$, we expect $D_{L,BN}/D_{L,BB} \sim 0.4 (m_{B}/12M_\odot)^{1/2}$ and $D_{L,NN}/D_{L,BB} \sim 0.2 (m_{N}/1.3M_\odot)^{1/2}$. Thus we find the LIGO BB volume is $\times 20(250)$ larger than the LIGO BN (NN) volume, assuming same characteristic masses for the BH and NN in all mergers. Thus, for models R1a-R1c $N_{L,BN}/N_{L,BB} \sim (1/20)[0.12,0.44] \sim [0.006,0.02]$.  Whereas $N_{L,BN}/N_{L,BB} \sim [0.01,0.16]$ for models in Table~\ref{tab:r4bx100}. Similarly, $N_{L,NN}/N_{L,BB}<0.02$ for the models discussed here. At time of writing (end of O3a) and as reported by LIGO through its low-latency alert system \footnote{https://gracedb.ligo.org/superevents/public/O3/}\footnote{https://emfollow.docs.ligo.org/userguide/}, the ratio of BH-NS/BH-BH and NS-NS/BH-BH merger numbers reported $\approx 4/42(6/42)$.  (For this qualitative discussion, we  take the reported classifications without qualification and without attempt to propagate Poisson or other uncertainties, not addressing the prospects for false alarms or the interpretation of the classification boundaries adopted by LIGO's reporting, e.g., that $m<3M_\odot$ is denoted a neutron star.)  The BN/BB ratios are closer to that expected from a $M^{-2}$ IMF with little mass segregation and the observed NN/BB ratios are almost an order of magnitude larger than our upper limits. The latter result implies that even if AGN dominate the observed BH-BH detection rate, this channel cannot be the dominant contribution to the NS-NS mergers detected by LIGO. The range of the BH-X/BH-BH merger fraction, i.e. relative rate of mergers involving \emph{lower mass gap} objects should be similar to the BH-NS/BH-BH merger fraction in this channel. As the population of mergers grows in future LIGO/Virgo runs, we will be able to constrain the overall mass function and disk models in the AGN channel more strongly.

If a fraction $f_{AGN}$ of the observed rate of BB mergers (${\cal}{R}_{BB} \sim 100 \rm{Gpc}^{-3} \rm{yr}^{-1}$) come from AGN, we expect a rate of BN (NN) mergers ${\cal}{R}_{BN} \sim f_{AGN}[10,300]\rm{Gpc}^{-3} \rm{yr}^{-1}$,(${\cal}{R}_{NN} \sim f_{AGN}[0.2,400]\rm{Gpc}^{-3} \rm{yr}^{-1}$) from the same channel. Since we estimate the BN(NN) LIGO search volume is $\sim 1/20(1/250)$ of the BB LIGO search volume, we expect a LIGO-detectable rate of BN(NN) mergers from AGN of $ {\cal}{R}_{L,BN} \sim f_{AGN}[0.5,15]\rm{Gpc}^{-3} \rm{yr}^{-1}$, (${\cal}{R}_{L,NN} \sim f_{AGN}[10^{-3},1.6]\rm{Gpc}^{-3} \rm{yr}^{-1}$) respectively. 

Finally, we note that since this paper was submitted, LIGO has announced event GW190814 \citep{Abbott20} with a mass ratio of $q=0.1$. This event may have been a BH-NS merger or a BH-BH merger. In the context of the present results, a BH-NS $q=0.1$ merger would be the \emph{expectation} for an AGN-disk driven merger in a fairly typical galactic nucleus with mass function $\sim M^{-1.5}$, which might correspond to e.g. a Kroupa IMF allied with some mass segregation or top-heavy star formation. If instead GW190814 is a BH-BH merger, the lack of high spins suggest that this is a merger of two first generation BH and therefore not a migration trap merger, but rather a merger in the bulk disk. A $q=0.1$ merger is about a $1/20$ event in bulk disk mergers, so an AGN origin for this BH-BH merger would imply that about half of the mergers detected by LIGO in O3 are associated with AGN disks.

\subsection{Implications for EM observations}
A BH-BH merger in an AGN disk will generate a much lower luminosity prompt EM counterpart \citep{McK19b} than a BH-NS or BH-WD merger where the secondary is tidally disrupted, although see \citet{Graham20} for a super-luminous candidate counterpart. For the runs in Table~\ref{tab:r1bx100}, the fraction of BH-NS mergers with BH mass $<7M_{\odot}(<10M_{\odot})$ is approximately $\sim 0.11(0.25)$. Thus we expect fraction $f_{EM} \sim [0.1,0.2]$ of the BH-NS mergers in Table~\ref{tab:r1bx100} to yield EM counterparts from a TDE outside the BH event horizon. Such signatures may be muffled by the AGN disk, or they may show up as fast-rise, slower decay flares superimposed on the AGN lightcurve with relatively similar lightcurves.  

Therefore from the previous section, we expect ${\cal}{R}_{EM, BN} \approx (f_{EM}/0.1)f_{AGN}[1,30]\rm{Gpc}^{-3} \rm{yr}^{-1}$. Thus within redshift $z<1$, corresponding to a luminosity distance of $\sim 6.7$Gpc, we should expect 
\begin{equation}
    {\cal}{R}_{EM,BN}(z<1) \approx [30,900]{\rm yr}^{-1} \left(\frac{f_{EM}}{0.1}\right)\left(\frac{f_{AGN}}{0.1}\right)
\end{equation}
number of EM counterparts to BH-NS mergers in AGN disks. The curious high luminosity, fast-rise, slow-decay events seen in large AGN samples may correspond to such events \citep{Graham17,Cannizzaro19}.

\section{Conclusions}
We investigated the merger of different types of compact objects in the AGN channel in a series of Monte Carlo simulations spanning a range of disk models and nuclear stellar mass functions. We found median mass ratios of NS-BH mergers in AGN disks are  $\tilde{q}=0.07\pm 0.06(0.14\pm 0.07)$ for mass functions $M^{-1(-2)}$. The ratio of NS-BH/BH-BH mergers in this channel spans $\sim [0.1,3]$. The ratio of NS-NS/BH-BH mergers in the AGN channel span $\sim [10^{-3},4]$. Taking into account the ratio of  NS-NS/BH-BH LIGO search volumes, preliminary O3 results rule out this channel as a dominant contribution to NS-NS mergers.  The rate of occurrence of lower mass gap events like GW190425z in this channel depends strongly on the nuclear IMF and the level of mass segregation. Compact object merger ratios derived from LIGO can restrict models of nuclear mass functions, mass segregation and populations embedded in AGN disks. The recently reported event GW190814 with mass ratio $q=0.1$, if a BH-NS merger, would be an \emph{expectation} value for the AGN channel. If instead, GW190814 is a BH-BH merger, it would correspond to a 1 in 20 event in bulk disk mergers in the AGN channel, implying about half of the O3 BH-BH mergers are associated with AGN.

The expected number of EM counterparts to NS-BH mergers in AGN disks at $z<1$ is $\sim [30,900]{\rm{yr}}^{-1}(f_{AGN}/0.1)$.  EM searches for flaring due to embedded events in large AGN surveys will complement LIGO constraints on AGN models and their embedded populations.

{\section{Acknowledgements.}} BM \& KESF are supported by NSF grant AST-1831412 and Simons Foundation award 533845. ROS is supported by NSF PHY-1707965, AST-1909534, and PHY-1912632. Thanks to the referee for a useful report that helped strengthen our conclusions. 

\section{Data Availability} The data used here are available on reasonable request from corresponding author, B. McKernan.


\begin{thebibliography}{99}
\bibitem[\protect\citeauthoryear{Aasi et al.}{2015}]{Aasi15} Aasi J. et al., 2015, Classical \& Quantum Gravity, 32, 4001
\bibitem[\protect\citeauthoryear{Abbott et al.}{2019b}]{Abbott18Pop} Abbott B.P. et al., 2019a, ApJ, 882, L24
\bibitem[\protect\citeauthoryear{Abbott et al.}{2019a}]{Abbott18GWTC1} Abbott B.P. et al., 2019b, PhRvX, 9, 1040
\bibitem[\protect\citeauthoryear{Abbott et al.}{2020}]{Abbott20} Abbott B.P. et al., 2020, ApJ, 896, L44
\bibitem[\protect\citeauthoryear{Acernese et al.}{2015}]{Acernese15} Acernese F. et al., 2015, Classical \& Quantum Gravity, 32, 024001
\bibitem[\protect\citeauthoryear{Antonini}{2014}]{Antonini14} Antonini F., 2014, ApJ, 794, 106
\bibitem[\protect\citeauthoryear{Alexander \& Hopman}{2009}]{Alexander09} Alexander T. \& Hopman C., 2009, ApJ, 797, 1861
\bibitem[\protect\citeauthoryear{Bartko et al.}{2010}]{Bartko10} Bartko H. et al., 2010, ApJ, 708, 834
\bibitem[\protect\citeauthoryear{Bartos et al.}{2017}]{Bartos17} Bartos I. et al., 2017, ApJ, 835, 165
\bibitem[\protect\citeauthoryear{Baruteau et al.}{2011}]{Baruteau11} Baruteau C., Cuadra J. \& Lin D.N.C., 2011, ApJ, 726, 28
\bibitem[\protect\citeauthoryear{Baumgardt et al.}{2019}]{Baumgardt19} Baumgardt H., Amaro-Seoane P. \& Sch\"{o}del R., 2019, A\&A, 609, 28
\bibitem[\protect\citeauthoryear{Bellovary et al.}{2016}]{Bellovary16} Bellovary J. et al., 2016, ApJ, 819, L17
\bibitem[\protect\citeauthoryear{Bogdanovic et al.}{2007}]{Bogdanovic07} Bogdanovic T., Reynolds C.S. \& Miller M.C., 2007, ApJ, 661, L147
\bibitem[\protect\citeauthoryear{Cannizzaro et al.}{2020}]{Cannizzaro19} Cannizzaro G. et al., 2020, MNRAS (accepted), arXiv:2001.07446
\bibitem[\protect\citeauthoryear{Chatziioannou et al.}{2019}]{KC2019} Chatziioannou K. et al., 2019, PhRvD, 100, 104015
\bibitem[\protect\citeauthoryear{Dittmann \& Miller}{2020}]{Dittmann20} Dittmann A. \& Miller M.C., 2020, MNRAS, 493, 3732
\bibitem[\protect\citeauthoryear{Ford \& McKernan}{2019}]{Ford19} Ford, K.E.S. \& McKernan B., 2019, MNRAS, 490, L42
\bibitem[\protect\citeauthoryear{Fishbach et al.}{2017}]{Fishbach17} Fishbach M., Holz D.E. \& Farr B., 2017, ApJ, 840, L24
\bibitem[\protect\citeauthoryear{Generozov et al.}{2018}]{Generozov18} Generozov A., Stone N.C., Metzger B.D. \& Ostriker J.P., 2018, MNRAS, 478, 4030
\bibitem[\protect\citeauthoryear{Gerosa \& Berti}{2019}]{Gerosa19} Gerosa D. \& Berti E., 2019, PhRvD, 100, 1301
\bibitem[\protect\citeauthoryear{Graham \& Spitler}{2009}]{Graham09} Graham A.W. \& Spitler L.R., 2009, MNRAS, 397, 2148
\bibitem[\protect\citeauthoryear{Graham et al.}{2017}]{Graham17} Graham M.J. et al., 2017, MNRAS, 470, 4112
\bibitem[\protect\citeauthoryear{Graham et al.}{2020}]{Graham20} Graham M.J. et al., 2020, PhRvL, 124 ,1102 
\bibitem[\protect\citeauthoryear{Hailey et al.}{2018}]{Hailey18} Hailey C.J. et al., 2018, Nature, 556, 70
\bibitem[\protect\citeauthoryear{LIGO}{2019}]{LIGO18} LIGO \& VIRGO Scientific Collaborations, 2019a, ApJ, 882, L24 
\bibitem[\protect\citeauthoryear{LIGO}{2020}]{LIGO20} LIGO \& VIRGO Scientific Collaborations, 2020, ApJ (submitted), arXiv:2001.01761
\bibitem[\protect\citeauthoryear{Leigh et al.}{2018}]{Leigh18} Leigh N.W.C. et al., 2018, MNRAS, 474, 5672
\bibitem[\protect\citeauthoryear{Lu et al.}{2013}]{Lu13} Lu J. et al., 2013, ApJ, 764, 155
\bibitem[\protect\citeauthoryear{McKernan et al.}{2012}]{McK12} McKernan B. et al., 2012, MNRAS, 425, 460

\bibitem[\protect\citeauthoryear{McKernan et al.}{2014}]{McK14} McKernan B. et al., 2014, MNRAS, 441, 900
\bibitem[\protect\citeauthoryear{McKernan et al.}{2018}]{McK18} McKernan B. et al., 2018, ApJ, 866, 66
\bibitem[\protect\citeauthoryear{McKernan et al.}{2020}]{McK19a} McKernan B. Ford K.E.S., O'Shaughnessy R. \& Wysocki D., 2020, MNRAS, 494, 1203
\bibitem[\protect\citeauthoryear{McKernan et al.}{2019b}]{McK19b} McKernan B. et al., 2019, ApJL, 884, L50
\bibitem[\protect\citeauthoryear{Miralda-Escud\'{e} \& Gould}{2000}]{Miralda00} Miralda-Escud\'{e} J. \& Gould A., 2000, ApJ, 545, 847
\bibitem[\protect\citeauthoryear{Morris}{1993}]{Morris93} Morris M., 1993, ApJ, 403, 496
\bibitem[\protect\citeauthoryear{Paaredekooper et al.}{2010}]{Paar10} Paardekooper S. -J., Baruteau C., Crida A. \& Kley W., 2010, MNRAS, 401, 1950
\bibitem[\protect\citeauthoryear{Schodel et al.}{2018}]{Schodel18} Sch\"{o}del R. et al., 2018, A\&A, 609, 27
\bibitem[\protect\citeauthoryear{Secunda et al.}{2019}]{Secunda18} Secunda A. et al., 2019, ApJ, 878, 85
\bibitem[\protect\citeauthoryear{Secunda et al.}{2020}]{Secunda20} Secunda A. et al., 2019, ApJ, submitted (arXiv:2004.11936)
\bibitem[\protect\citeauthoryear{Sigurdsson \& Phinney}{1993}]{SigPhin} Sigurdsson S. \& Phinney E.S., 1993, ApJ, 415, 631
\bibitem[\protect\citeauthoryear{Sirko \& Goodman}{2003}]{SG03} Sirko E. \& Goodman J., 2003, MNRAS, 341, 501

\bibitem[\protect\citeauthoryear{Stone et al.}{2017}]{Stone17} Stone N.C. et al., 2017, ApJ, 464, 946
\bibitem[\protect\citeauthoryear{Subr \& Karas}{2005}]{Subr05} Subr L. \& Karas V., 2005, A\&A, 433, 405
\bibitem[\protect\citeauthoryear{Syer et al.}{1991}]{Syer91} Syer D., Clarke C. \& Rees M.J., 1991, MNRAS, 250, 505
\bibitem[\protect\citeauthoryear{Tanaka et al.}{2002}]{Tanaka02} Tanaka H., Takeuchi T. \& Ward, W.R., 2002, ApJ, 565, 1257 
\bibitem[\protect\citeauthoryear{Tichy \& Marronetti}{2008}]{Tichy08} Tichy W. \& Marronetti P., 2008, Phys. Rev. D, 78, 1501
\bibitem[\protect\citeauthoryear{Thompson et al.}{2005}]{Thompson05} Thompson T.A., Quataert E. \& Murray N., 2005, ApJ, 630, 167
\bibitem[\protect\citeauthoryear{Udall et al.}{2019}]{Udall19} Udall R. et al., 2019, ApJ (submitted), arXiv:1912.10533
\bibitem[\protect\citeauthoryear{Varma et al.}{2019}]{Varma19} Varma V., Gerosa D., Stein L., Hebert F., Zhang H., 2019, Phys. Rev. Lett., 122 1101
\bibitem[\protect\citeauthoryear{Woosley}{2017}]{Woosley17} Woosley S.E., 2017, ApJ, 836, 244
\bibitem[\protect\citeauthoryear{Wysocki et al.}{2018}]{Wysocki18} Wysocki  D., Lange J., \& O'Shaughnessy R., 2018, arxiv:1805.06442
\bibitem[\protect\citeauthoryear{Yang et al.}{2019}]{Yang19} Yang Y. et al. 2019, PhRvL, 123, 1101
\bibitem[\protect\citeauthoryear{Zackay et al.}{2019}]{Zackay19} Zackay B. et al. 2019, PhRvD (submitted), arXiv:1910.09528

\end{thebibliography}
\end{document}